\documentclass[conference]{IEEEtran}

\usepackage{cite}
\usepackage{amsmath,amssymb,amsfonts}
\usepackage{algorithmic}
\usepackage{graphicx}
\usepackage{textcomp}
\usepackage{xcolor}
\usepackage{url}

\usepackage[varg]{txfonts}   

\usepackage{graphicx}
\usepackage{xspace}
\usepackage{xcolor}
\usepackage{colortbl}

\ifCLASSOPTIONcompsoc
    \usepackage[caption=false, font=normalsize, labelfont=sf, textfont=sf]{subfig}
\else
\usepackage[caption=false, font=footnotesize]{subfig}
\fi

\newcommand{\hpc}{HPC\xspace}

\newcommand{\nexa}{NET4EXA\xspace}
\newcommand{\ai}{AI\xspace}

\begin{document}
\title{\nexa: Pioneering the Future of Interconnects for Supercomputing and AI}

%
%
\author{\IEEEauthorblockN{
Michele Martinelli\IEEEauthorrefmark{5},
Roberto Ammendola\IEEEauthorrefmark{6},
Andrea Biagioni\IEEEauthorrefmark{5},
Carlotta Chiarini\IEEEauthorrefmark{5}\IEEEauthorrefmark{9},
Ottorino Frezza\IEEEauthorrefmark{5},\\
Francesca Lo~Cicero\IEEEauthorrefmark{5},
Alessandro Lonardo\IEEEauthorrefmark{5},
Pier Stanislao Paolucci\IEEEauthorrefmark{5},\\
Elena Pastorelli\IEEEauthorrefmark{5},
Pierpaolo Perticaroli\IEEEauthorrefmark{5},
Luca Pontisso\IEEEauthorrefmark{5},
Cristian Rossi\IEEEauthorrefmark{5}\IEEEauthorrefmark{9},\\
Francesco Simula\IEEEauthorrefmark{5},
Piero Vicini\IEEEauthorrefmark{5}, \\ \\
David Colin\IEEEauthorrefmark{1},
Grégoire Pichon\IEEEauthorrefmark{1},
Alexandre Louvet\IEEEauthorrefmark{1},\\
John Gliksberg\IEEEauthorrefmark{1},
Claire Chen\IEEEauthorrefmark{1}, \\ \\
Matteo Turisini\IEEEauthorrefmark{2},
Andrea Monterubbiano\IEEEauthorrefmark{2}, \\ \\
Jean-Philippe Nominé\IEEEauthorrefmark{3},
Denis Dutoit\IEEEauthorrefmark{3},
Hugo Taboada\IEEEauthorrefmark{3},
Lilia Zaourar\IEEEauthorrefmark{3}, 
Mohamed Benazouz\IEEEauthorrefmark{3}, \\ \\
Angelos Bilas\IEEEauthorrefmark{4},
Fabien Chaix\IEEEauthorrefmark{4},
Manolis Katevenis\IEEEauthorrefmark{4},
Nikolaos Chrysos\IEEEauthorrefmark{4},
Evangelos Mageiropoulos\IEEEauthorrefmark{4},
Christos Kozanitis\IEEEauthorrefmark{4}, \\ \\
Thomas Moen\IEEEauthorrefmark{7},
Steffen Persvold\IEEEauthorrefmark{7},
Einar Rustad\IEEEauthorrefmark{7},\\ \\
Sandro Fiore\IEEEauthorrefmark{8},
Fabrizio Granelli\IEEEauthorrefmark{8},
Simone Pezzuto\IEEEauthorrefmark{8}, 
Raffaello Potestio\IEEEauthorrefmark{8},\\
Luca Tubiana\IEEEauthorrefmark{8},
Philippe Velha\IEEEauthorrefmark{8},
Flavio Vella\IEEEauthorrefmark{8}, \\ \\
Daniele De Sensi\IEEEauthorrefmark{9},
Salvatore Pontarelli\IEEEauthorrefmark{9}, \\ \\
}

\IEEEauthorblockA{\IEEEauthorrefmark{1}Atos - Eviden (BULL), France [name].[surname]@eviden.com}
\IEEEauthorblockA{\IEEEauthorrefmark{2}CINECA, Italy [m.turisini, a.monterubbiano]@cineca.it}
\IEEEauthorblockA{\IEEEauthorrefmark{3}Commissariat à l'énergie atomique et aux énergies alternatives (CEA), France [name].[surname]@cea.fr}
\IEEEauthorblockA{\IEEEauthorrefmark{4}Foundation for Research and Technology - Hellas (FORTH), Greece [name].[surname]@ics.forth.gr}
\IEEEauthorblockA{\IEEEauthorrefmark{5}INFN, Sezione di Roma, Italy [name].[surname]@roma1.infn.it}
\IEEEauthorblockA{\IEEEauthorrefmark{6}INFN, Sezione di Roma Tor Vergata, Italy [name].[surname]@roma2.infn.it}
\IEEEauthorblockA{\IEEEauthorrefmark{7}Numascale, Norway [name].[surname]@numascale.com}
\IEEEauthorblockA{\IEEEauthorrefmark{8}Università di Trento, Italy [name].[surname]@unitn.it}
\IEEEauthorblockA{\IEEEauthorrefmark{9}Università "Sapienza" di Roma, Italy [name].[surname]@uniroma1.it}
}

\maketitle


\section{Introduction} 
High Performance Computing (\hpc) is a complex domain where several cutting-edge technologies must interoperate to provide the most efficient environment for Supercomputing applications and Artificial Intelligence (\ai), like - but not limited to - large language models. Main technologies involved are computing systems (processors, accelerators), storage systems (non-volatile memory, filesystem, archiving) and network systems (management and interconnection networks). The interconnection network, in particular, is a critical technology since it has to ensure low-latency and high-bandwidth communications between the thousands of servers that are used for \hpc simulations and \ai models training. Furthermore, the role of the interconnect is becoming more important as future \hpc and \ai applications require scaling to extreme processor counts and dataset sizes.
The interconnect technology is currently dominated by US-based actors: NVidia with InfiniBand, HPE with Slingshot, Broadcom with Ethernet. One of the major outcomes of the \nexa project is to deliver a competitive European Interconnection network, namely BXIv3, ensuring sovereignty in this technology for the next decade. This is vital for providing independent technology sourcing and supplying critical industries and public sectors in Europe.

In this perspective, a number of technology components are tackled by \nexa for the BXIv3 development:

\textbf{Exaflops Computing Scale:} while the previous BXI generation, BXIv2, can connect 64k nodes, BXIv3 will scale to 8M interconnected endpoints, supporting larger clusters and flexible data center organization. It enables easier integration and coordination of specialized clusters, enhancing performance and flexibility.

\textbf{Energy-Efficient Interconnect:} focuses on energy efficiency by integrating with GPU technology, supporting liquid cooling and offering power consumption monitoring. The project aims to meet the 20 MW power limit for Exascale clusters while optimizing energy use in data centers.

\textbf{Reliable Interconnect:} includes features like error correction, end-to-end re-transmission and adaptive routing to ensure high network reliability, minimizing downtime and reducing operational costs. The target is a 62,000-hour Mean Time Before Failure (MTBF) for hardware.

\textbf{Secure Interconnect:} implements multiple security layers, including safe system configuration, traffic isolation, job-specific access control and encryption of traffic payloads, ensuring secure communication in sensitive environments.

\textbf{Interoperable Interconnect:} uses Ethernet-based communication for easy integration with existing datacenter networks. It supports seamless interaction with backbone and management networks, reducing hardware costs and avoiding performance bottlenecks.

\subsection{Interconnection Network architecture}
Interconnection network forms the backbone of supercomputers and datacenters: it allows the many thousands of compute and management servers, vector and GPU accelerators, storage and I/O subsystems to communicate and thus cooperate with each other. The throughput of accelerators, memory hierarchies, compute servers and storage keeps increasing at a high pace, and at the same time fine-grain parallelism as well as Non-volatile memory (NVM) storage devices demand lower latency. These are the motivations of the \textit{Call for Innovation Action in Low Latency and High Bandwidth Interconnects} to which the \nexa proposal was submitted~\cite{NET4EXA_site,NET4EXA_call}, and these same motivations stand for the BXI roadmap evolution since a decade ago and further into the current decade.

\begin{figure}[h!]
\centering
\includegraphics[width=0.75\linewidth]{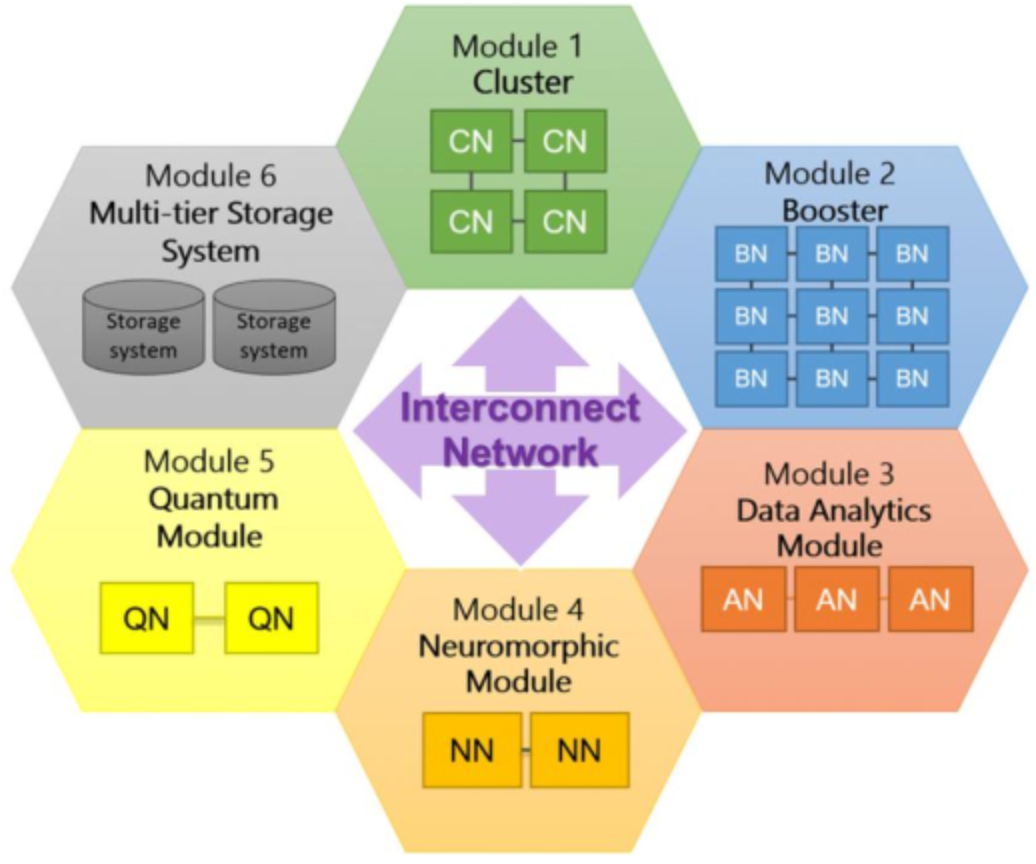}
\caption{\label{fig:MSA}Example of Modular Supercomputer Architecture (MSA).}
\end{figure}

BXI (Bull eXascale Interconnect)\cite{BXI_PAPER} is a European interconnect developed \& industrialized by BULL. It is made of two hardware components: the Network Interface Card (NIC) that allows the servers to access to the network, and the switch that establishes communication between the servers. BXI has been designed to support modern modular and heterogeneous computing centers, inspired by the Modular Supercomputer Architecture (MSA)\cite{suarez2019msa} (see \figurename\ref{fig:MSA}): the network backbone, federating specialized clusters, will present the supercomputer as an aggregation of resources that are organized to facilitate the mapping of application workflows, in accordance with the convergence of \hpc, Datacentre and High-Performance Data Analytics (HPDA) and Artificial Intelligence workloads.

Version 2 of BXI, BXIv2, has been deployed at CEA in Exa1-HF BullSequana XH2000, ranked 14th in November 2021 Top500 list. It provided 100Gb/s connectivity to servers and competitive low latency, and scaled up to 64k nodes. 
To further support such as modular architectures, the new BXI version, BXIv3, implemented in this project, will use Ethernet as its underlying link technology, as opposed to the proprietary link protocol of the previous BXIv2. In this way BXIv3 will provide native support for both traditional IP/Ethernet traffic as well as for the high-performance BXI/Portals traffic, and for the federation of clusters through L3 IP routers. Scalability to large system sizes will also be ensured through increased NIC capacities.
In comparison with BXIv2, \nexa project main deliverable BXIv3, provides 4 to 8 times higher throughput, 10 times higher message rate, 2 times lower latency, Ethernet link technology and Internet Protocol support for interoperability, multi-fabric federation, congestion management, message fragmentation, security improvements, and several other new features.

To provide high performance and reduce energy consumption, modern HPC network interfaces must eliminate operating system calls and redundant message copies while sending and receiving messages, i.e. support user-level, zero-copy communication. The BXI NIC does so by providing hardware acceleration for Send/Receive (two-sided) message passing semantics, by providing Remote DMA (RDMA) for one-sided operations, and through hardware support for collective operations. User processes issue their commands using virtual-address arguments, and the NIC translates them into physical addresses. The BXI hardware allows for efficient implementations of the Portals Application Programming Interface (API)\cite{portals_site}.

Quality-of-service, and particularly the management of congestion, is an important and increasing concern in modern HPC/HPDA/AI systems or datacentres, with their heterogeneous compute nodes (GPUs, AI accelerators and storage). Platforms are shared among users and applications with different demands and the isolation between the jobs is required. With increased scale and increased use of communication, the traffic patterns may well become unpredictable and cause congestion. The latency of critical messages, such as synchronization barriers, must stay unaffected. Moreover, the applications based on small messages at high throughput coexists with nodes generating larger frames. The fabric must ensure a uniform repetitive deterministic behavior across all applications. Besides, to optimize cost by reducing the number and the rate of the links, a higher utilization ratio is desirable.

\subsection{Project roadmap}
\begin{figure}[h!]
\centering
\includegraphics[width=0.9\linewidth]{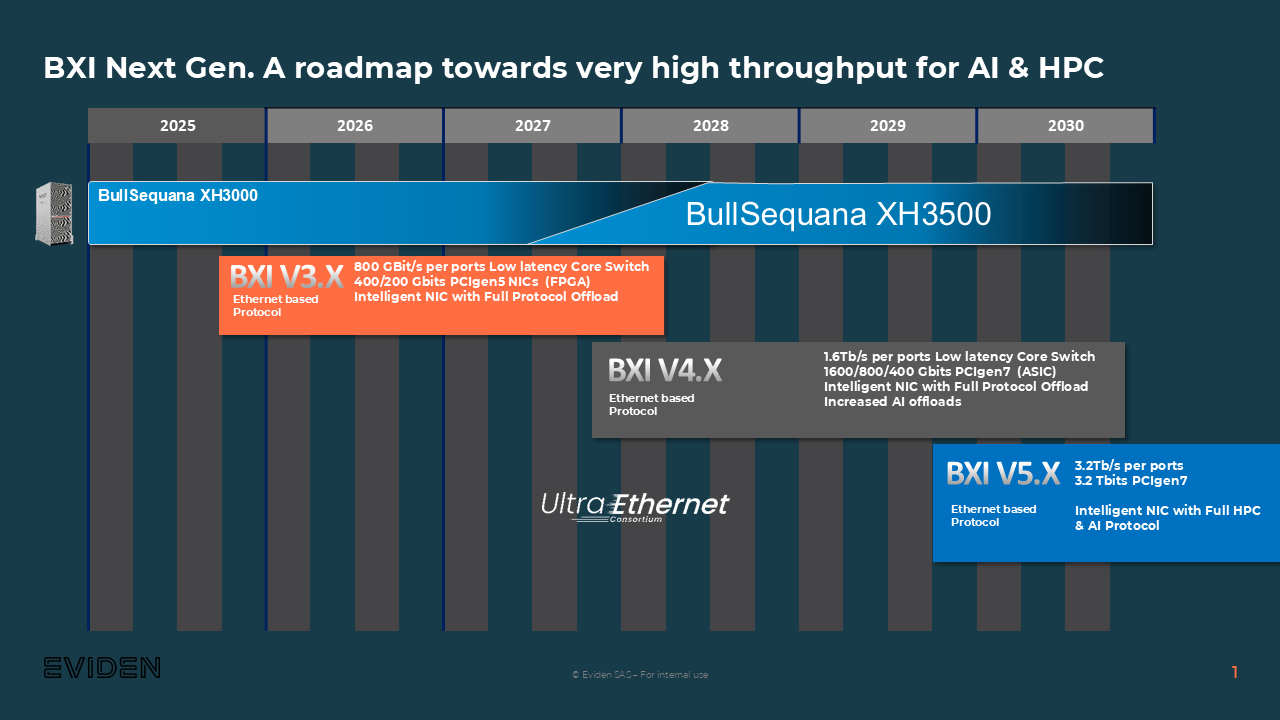}

\caption{\label{fig:roadmap}BXI Next Generation roadmap}
\end{figure}

Throughout the evolution of the BXI (“Bull eXascale Interconnect”) roadmap, from BXIv1 in 2017 and BXIv2 in 2021 to BXIv3 planned for 2025\---26 and beyond, the teams in charge carefully monitored technology and market evolution and defined a roadmap to maintain competitiveness at a global scale. \figurename \ref{fig:roadmap} provides an overview of the current roadmap.

Our objectives and \textit{Key Performance Indicators} (KPIs) relate to the BXIv3 Switch, the BXIv3 NIC also known as \textit{NICIA}, and the system software required for supporting HPC and AI applications over them; at places, we also refer to the preparatory work to be carried out in this project for the next step in the roadmap, BXIv4. To meet its ambitious goals related to technical features, schedule and budget, the project will leverage a commercially available switch chip from a third party, extended with IP specified and designed by BULL, and an FPGA-based NIC. 

NET4EXA chooses to use flexible FPGA technology for the BXIv3 NIC, to allow step-by-step development and verification; the resulting verified and robust design will be ported into Application-Specific Integrated Circuit (ASIC) technology for the BXIv4 NIC, to reduce cost and energy consumption, when the NIC will move to mass production after NET4EXA fulfills its goals.

The deliverables of this project include:
\begin{itemize}
    \item the Pilot Testbed integrating BXIv3; 
    \item the Software that will run on it; 
    \item the electronic equipment that will host the NIC FPGA and the Switch ASIC; 
    \item the Configuration bitfile of the FPGA based NIC. 
\end{itemize}
As for the Switch ASIC, this is being taped-out and manufactured in the form of a commercial part by a third party, however with a custom and proprietary configuration, where the configuration has been specified and designed by and is the property of BULL.

\subsection{Objectives}

The objectives are grouped in four broad categories: \textit{performance}, \textit{scalability}, \textit{usability}, and
\textit{commerciality}. 
We then discuss how these objectives relate to the EuroHPC Work Programme, how they relate to the state-of-the-art, and the fact that we will achieve Technology Readiness Level (TRL) 8.
\subsubsection{Performance}
NET4EXA aims to achieve high-performance (peak throughput and low latency) levels and enhanced Quality of Service (QoS) across three key objectives.

The third-generation switch (v3) will support 64 ports, each delivering 8 times the performance of previous generation, with the flexibility to be reconfigured into two ports or further split into four ports. On the other hand, the switch will maintain about 200~nanosecond per-hop latency for small packets under low load conditions, comparable to other high-performance networks such as Slingshot~\cite{slingshot}.
Similarly, the third-generation Network Interface Card (NIC~v3) will provide a network-side bandwidth 4 times faster than previous generation, also configurable as two independent connections, and a 16-lane PCIe Gen5 interface on the host side. 
Looking ahead, the project will lay the groundwork for BXIv4, which is expected to double the throughput of BXIv3 while integrating PCIe Gen6 with Compute Express Link (CXL) support.


To ensure traffic isolation and efficient resource allocation, the NIC will incorporate eight Virtual Channels (VCs). 
Additionally, the switches will feature sixteen VCs, enabling adaptive routing and preventing deadlocks. Many of the scalability features implemented will also directly enhance QoS.


The BXIv3 NIC will feature a full hardware implementation of the Portals 3 interface over Ethernet, 
leveraging on embedded ARM CPUs to manage Portals process initialization and termination and offloading many and frequent software operations from the host to specialized
hardware on the NIC.

Unlike BXIv2, where Ethernet frames were encapsulated in Portals messages, BXIv3 will provide native TCP/IP v4 support over its Ethernet links and switches, also providing a set of TCP offloading features such as checksum offload and hardware packet segmentation/aggregation.  




Finally, in addition to virtual-to-physical address translation, the BXIv3 NIC will facilitate direct interactions with accelerators such as GPUs from NVIDIA, AMD, and Intel. This support will enable zero-copy data transfers, allowing direct access to GPU memory buffers and minimizing the need for redundant data copies.

\subsubsection{Scalability}
NET4EXA aims to achieve the scalability levels required by modern and future HPC and AI applications, supporting up to 64k nodes (NICs) per Cluster and up to 128 federated clusters of a Modular Supercomputer Architecture (MSA), shaping real-life Fat-Trees and/or Dragonfly+ (DF+) topologies. 
The routing is fixed within a conversation --- i.e. the combination of a process and an Message Passing Interface (MPI) communicator --- for the in-order delivery of messages. Minimal or non-minimal adaptive routing is guaranteed for out-of-order tolerant packets.
Further, we design mechanisms of traffic throttling based on notifications from network management, destination nodes, or switches allowing the option 
to drop stalled packets to manage congested scenarios.

Offload capabilities allow for memory access and instruction per message reduction increasing the energy efficiency. This objective guides the adoption of GPUs providing lower Watt/Flop ratio than general purpose processors and the support of specific technologies (like NVIDIA's GPUDirect) to prevent recopying of data into host memory.

The NIC mezzanine and the Switch boards will be liquid cooled offering better Power Usage Effectiveness (PUE) than air cooling.
Monitoring of power consumption of switches is essential and custom mechanisms to power-down selected links and/or entire switches are planned.
Finally,  the Silicon Photonics, prepared for BXIv4, will also reduce energy consumption.

Large-scale systems need  optimized mechanisms to ensure reliability.
The BXIv3 link-layer protocol will include error detection/correction, per-packet acknowledgment, and link-level retransmission, while the transport layer ensures retransmission of packets upon CRC error and timeout.
The above-mentioned adaptive routing allows for a significant reliability improvement, moving packets around failed links.
Finally, telemetry and network monitoring, including link-endpoint failure detection, will constantly provide statistics to the network management software.

\subsubsection{Usability}

BXIv3 will operate over Ethernet physical link protocol. The objectives for BXIv3 listed in this section are similar as the foreseen features of Ultra Ethernet~\cite{uec}, customized for HPC and AI workloads. 
\nexa will provide the required support in the Linux network stack to use BXIv3 transparently as an Ethernet NIC, offering transparent support for legacy network protocols (e.g. IP, TCP, UDP).
However, the full Ultra Ethernet compatibility is only planned for BXIv4: \nexa will pave the road for such full compatibility, in parallel with the ongoing standardization process by Ultra Ethernet Consortium (UEC).
The BXIv3  adopts PCIe Gen5 NIC-to-host processor interface. CXL cache coherence over this PCIe interface will be prepared during \nexa for implementation in the next roadmap version, BXIv4.

Like previous BXI versions, the NIC will implement the Portals 4 API (Application Programming Interface). 
\nexa will offer an efficient implementation of the libraries required for using the hardware features of BXIv3 for Portals.
Further, \nexa aims to provide a rich software ecosystem, by adapting existing and developing new software components for BXIv3, allowing the
deployment and use of BXIv3 in HPC and AI applications. 

Lastly \nexa will provide all required network management tools to install, configure, monitor and analyze the operation of BXIv3.

\subsubsection{Commerciality}

To achieve the desired TRL 8, \nexa starts with IP designs and preliminary implementations from previous projects (mostly RED-SEA) on a flexible platform (reconfigurable FPGA development boards). 
Small-scale testbeds will be installed since the very beginning of the project, and FPGA configurations and software will be updated on a regular basis as the project develops new IP. 
Further a larger testbed, our “Pilot”, will be built in a central
location and used remotely by all partners during the second half of the project. 

By the end of the project, BXIv3 will be implemented in our Pilot in the form of a system that will be complete and qualified, thus reaching TRL 8. 
Qualification will be achieved through the use cases.
Our list of real-life scientific applications and benchmarks includes Top500 and Graph500, large scale Brain Simulation, GROMACS (molecular dynamics), Quantum Espresso (electronic structure and material modelling), SPECFEM3D (seismic waves propagation), BERT (natural language processing), computing kernels implementing distributed-memory algorithms in different domains (scientific Computing, AI and ML data analytics).

\subsection{State of the art}
In the area of supercomputing interconnects, NVIDIA is the market leader, following its acquisition of Mellanox, with the InfiniBand (IB)~\cite{ib} –OFED compliant– protocols and line of products, and a large installed basis. IB provides high bandwidth and low latency; it is considered a high-cost technology relative to commodity Ethernet, since it addresses a narrower market, and it provides improved performance. 

Slingshot by Hewlett Packard Enterprise (HPE) is another major high-end interconnect technology~\cite{slingshot}. 

Cornelis Networks, derived from Intel OmniPath, is also a provider of high-end interconnects. 

All the above should be examined relative to Ethernet with the RDMA over Converged Ethernet (RoCE) extension, which encapsulates InfiniBand (IB) transport packets over Ethernet. Because it is based on the globally used and leading Ethernet technology, it is considered more mainstream than IB. However, Ethernet was not designed for HPC or AI at scale, and RoCE inherits this deficiency. This is precisely the reason why the Ultra Ethernet Consortium was founded, namely, to overcome these deficiencies and to deliver a standard and open interconnection technology for HPC and AI, thus competing with IB while at the same time being more promising than IB owing to its openness.

The BXI roadmap has started its alignment with Ethernet since several years ago, and BULL is now a founding and board member of Ultra Ethernet Consortium (UEC), committed to be strongly UEC-oriented in BXIv3 and fully compliant in BXIv4. Thus, with the work proposed here, we are well positioned to become one of the major global providers of HPC \& AI interconnects soon –actually, the only European player in this market at this moment.


The \nexa project  does not start from scratch; rather, it capitalizes on a solid foundation laid by several earlier European projects that have contributed valuable technological advances, architectural insights and implementation experience that have significantly influenced the current design and objectives.  

In particular, one of the key projects that has laid the groundwork for \nexa is Network Solution for Exascale Architectures (RED-SEA)\cite{redsea_micpro24} that contributed to the development of the new generation of European Interconnect, capable of powering the Exascale systems to come, through an economically viable and technologically efficient solution, leveraging interconnect technology (BXI) associated with standard and mature technology (Ethernet). RED-SEA did most of the preparatory work for BXIv3. \nexa will be its continuation, implementing BXIv3 and preparing the subsequent version, BXIv4.

DEEP Software for Exascale Architectures (DEEP-SEA) \cite{DEEPSEA_site} focus on the programming environment and software stack  for future European exascale systems by adapting all levels of the software stack to support highly heterogeneous compute and memory configurations and to allow code optimization like MPI hierarchical collectives (intra-node) and Remote DMA notifications.
BXIv2 has been used to connect 4 nodes of the DEEP System, which are part of the
testbed for DEEP-SEA. This enables the DEEP-SEA project's software stack to operate
seamlessly on these nodes.

European Pilot for ExaScale (EUPEX)\cite{EUPEX_site} aims to design, build and validate the first EU
platform for HPC, covering end-to-end the spectrum of required technologies with
European assets. The EUPEX prototype will be open, scalable and flexible, including
the modular OpenSequana compliant platform and the corresponding HPC software
ecosystem for the Modular Supercomputing Architecture.

The following results from EUPEX will be used in NET4EXA:
\begin{itemize}
    \item MPI hierarchical collectives (inter-node, ARM platform);
    \item Storage services (Key-Value store);
    \item Rhea processor boards (in the NET4EXA Pilot).
\end{itemize}

On the other hand, BXIv3, developed in NET4EXA, will be used as the inter-node interconnect in the
EUPEX pilot platform.

European Processor Initiative (EPI)\cite{EPI_site} is a collaborative project aimed at developing high-performance, energy-efficient processors to support the European Union's strategic autonomy in advanced computing. The \nexa Pilot testbed will include some nodes equipped with Rhea processors from EPI, to assess the integration of BXIv3 with EuroHPC computing technology.

\section{Project organization}
The \nexa project is structured around five key pillars each addressing specific objectives:
\begin{itemize}
    \item Architecture \& post-Exascale preparation (WP1) – aiming at defining BXI roadmap, assessing BXI compliance with emerging standards, protocols and chiplet interfaces, proposing ASIC integration scheme and finally delivering photonic transceivers as a first step towards a photonic roadmap;
    \item Communication Performance \& Offload (WP2) – developing BXIv3 hardware \& low-level software while improving BXI performance, security and energy efficiency;
    \item Fabric Management \& Scalability (WP3) – ensuring efficient, reliable, secure fabric management and offering a high level of scalability of the interconnect network;
    \item HPC and AI Software Ecosystem (WP4) – providing the essential components to achieve extensive coverage and superior performance over BXIv3 by enabling a rich software ecosystem
    \item Testbeds and Pilot Integration for Development and Evaluation (WP5) – focusing on system integration to evaluate the advancements in both the development of BXIv3 and BXIv4 interconnect technology.
\end{itemize}

\begin{figure}[h!]
\centering
\includegraphics[width=0.75\linewidth]{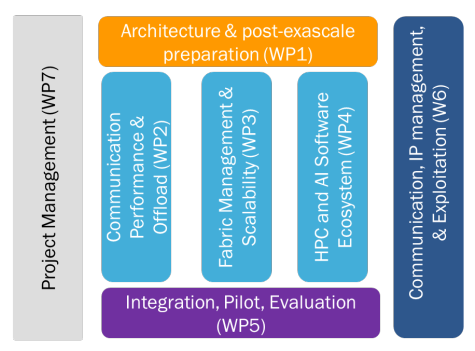}
\caption{\label{fig:img_wp}NET4EXA work-packages.}
\end{figure}

Around these pillars, NET4EXA defines seven work packages (five technical Work Packages (WP1 - WP5) and two other non-technical Work Packages: (WP6, WP7)) aimed at achieving the project's defined objectives, shown in \figurename \ref{fig:img_wp} and detailed in the next subsections.

\subsection{Architecture \& post-exascale preparation (WP1)}
The consortium will establish BXI requirements roadmap (Task 1.1) to steer the technical efforts concerning BXIv3 and the preparation work for next generations of interconnect and the main project objectives.
WP1 will also evaluate BXI compliance with emerging standards, specifically those from the Ultra Ethernet Consortium (Task 1.2)
The project will also explore architectural advancements for the next version of BXI (BXIv4) in Task 1.4. It involves the evaluation of chip architectures and optimizations to transition from FPGA-based implementations to a commercial ASIC chip. 
Finally, Task 1.5 aims to integrate silicon photonics solutions for BXIv3 and BXIv4, including the design and evaluation of photonic transceivers, chiplet-based co-packaged optics, and all-optical interconnects for future BXI versions. 

\subsection{Communication Performance \& Offload (WP2)}
WP2 will develop NICIA hardware alongside its low-level software to achieve high communication performance. 
The idea is to develop and deliver verified chip modules in hardware descriptive languages (RTL, Register Transfer Language) for the NIC hardware such that the combination of hardware and software can be validated together. RTL development will be done in both WP2 and WP3. The RTL integration and generation of FPGA bit-files will be done as part of WP2.
A dedicated task (Task 2.2) develops the low-level software for the network interface controller. This software, including device drivers, the Portals communication library and comprehensive test suites, is built in tandem with the NIC RTL releases. 

Additionally (Task 2.3) , the studies on encryption algorithm, on-NIC data processing and triggered operation extensions are also planned to address security at the network level and offloaded computation in NIC. It also examines in-network computing options, such as offloading collective operations to switches, and conducts studies on energy efficiency, comparing host-based versus NIC-offloaded communication and proposing power-aware algorithms.

Finally, the hardware and software parts will undergo bring-up and validation to support the foreseen developments outlined in this work package.

\subsection{Fabric Management \& Scalability (WP3)}
The project aims to ensure efficient, reliable, and secure network traffic routing within the fabric, in order to achieve an energy efficient interconnect network. 

It will also provide a mature and homogeneous fabric management software for BXIv3 switches (Task 3.1), including advanced routing algorithms for various topologies, topology discovery support and security features such as traffic isolation through VLAN/partition keys, to simplify the overall management process.

In parallel, Task 3.2 develops a comprehensive software stack for managing both NICs and switches, offering tools for installation, configuration, monitoring and diagnostics.

Moreover, it will secure smooth traffic flow through congestion management features by implementing and verifying advanced congestion management techniques at both the NIC and switch levels, with extensive testing on pilot testbeds and large-scale simulations, establishing a mature and homogeneous fabric management software stack facilitating high scalability  and inter-module communications (Task 3.3 and Task 3.4).

Finally, Task 3.5 is dedicated to enhancing energy efficiency by developing dynamic power reduction of network resources during idle periods, alongside power monitoring features for BXIv3 switches integrated into a Smart Energy Management Suite.

\subsection{HPC and AI Software Ecosystem (WP4)}
WP4 will provide the necessary software support for the interconnect and enable low-latency and high-bandwidth communication to HPC and AI applications, leveraging hardware offloading technologies introduced by WP2. WP4 will break down the software ecosystem into five areas:
\begin{enumerate}
    \item HPC programming models (Task 4.1), by developing a Unified BXI Communications Layer (UBCL) library to support high-level APIs for popular HPC and AI runtime systems (e.g. MPI). This effort involves adapting and integrating existing MPI implementations (e.g., OpenMPI and MPC), as well as extending support to Partitioned Global Address Space (PGAS) programming models and providing an Open Fabrics Interface (OFI) library to further exploit hardware optimizations. 
    \item Accelerators and AI (Task 4.2), by integrating BXIv3 within popular AI libraries (NCCL/RCCL) through a new Portals transport layer, enabling GPU-to-GPU communications across major GPU vendors, exploring GPU-triggered communications that allow kernels to initiate data transfers without CPU intervention. 
    \item Storage I/O (Task 4.3), by adding support for various storage abstractions and services over BXIv3.
    \item Datacentre application deployment (Task 4.4) concentrates on data center applications deployment by providing streamlined support for remote procedure calls (RPC) and virtualization. 
    \item Fabric-aware communication optimizations (Task 4.5) that exploit the inherent topological features of the system. This includes optimizing MPI rank placement based on network topology, as well as providing transparent support for multi-NIC configurations.
\end{enumerate}

\subsection{Testbeds and Pilot Integration for Development and Evaluation (WP5)}
WP5 will define and provide a wide array of application and service use cases to co-design and evaluate current and future BXI network interconnects. 

Task 5.1 is dedicated to procuring, integrating, and evaluating small-scale testbeds, which include a Bring-Up Testbed (BUT) for BXIv3 evaluation at BULL and multiple Features Development \& Research Testbeds (FDRT) deployed at CEA, FORTH, and INFN facilities. These testbeds, employing state-of-the-art FPGA platforms, GPUs, and RISC-V components, enable early-stage development and validation of the hardware and software components. Building on these efforts, Task 5.2 focuses on the procurement, integration, and large-scale demonstration of a Pilot testbed that brings together a heterogeneous mix of off-the-shelf and custom hardware—including CPUs, GPUs, and specialized accelerators—all hosted within a BULL rack and interconnected via BXIv3. This Pilot, managed collaboratively by BULL and CINECA with support from CEA, is intended for performance evaluation in production-like environments and may later be deployed in a Tier0 data center.

Parallel to the testbed deployments, Task 5.3 targets the porting, optimization, and co-design of selected HPC and AI application use-cases, as well as relevant service use-cases such as parallel filesystems and storage. 
Complementarily, Task 5.4 focuses on the development of BXIv3 boards—including NIC mezzanine/PCIe boards and switch boards—through successive stages of industrialization  and the integration of supporting hardware components such as mechanical and cooling systems, ensuring robust unitary testing and validation.
Task 5.5, on the other hand, is devoted to studies on the integration of BXI with CPUs and accelerators developed under other EuroHPC initiatives, with activities mainly devoted to interfacing with ARM and RISC-V processors via PCIe, or PCIe-based technologies, or via a direct connection to the host Network on Chip (NoC).

\subsection{Communication, IP management, \& Exploitation (WP6) and Project Management (WP7)}
In WP6 and WP7 the consortium's focus will be on Communication, IP management, and exploitation activities. This includes elevating the visibility of the NET4EXA project. Moreover, it will oversee the management and safeguarding of the IPs generated within the project. 


\section{Concluding remarks}
\nexa project will deliver a comprehensive interconnect solution composed of NIC and switch boards, transceivers, cables, and supporting software components. Two variants of the BXIv3 NIC board will be developed to address different segments of the market: a mezzanine version designed for high-density blade servers, such as those in the BullSequana XH3000 rack—targeting high-end HPC clusters—and a PCIe version intended for standard rack-mounted servers, thereby reaching a broader market and supporting OEM integration. Both NIC boards will incorporate FPGA technology, enabling incremental feature enhancements. However, 
a transition to an ASIC-based solution is planned for the next-generation BXIv4 NIC.

A single switch board will also be developed, deployable in either liquid-cooled or air-cooled enclosures, with the liquid-cooled variant specifically designed for integration into the BullSequana XH3000 infrastructure.

In parallel, \nexa will conduct analysis, design, prototyping and evaluation of advanced, innovative features to be eventually integrated in the next generation BXI interconnect to enhance the performances of the network architecture.

Beyond hardware and software development, the project is expected to stimulate the growth of related service activities, including consulting, support, maintenance and marketing. These efforts will contribute to job creation and offer a strategic opportunity for the consortium partners and broader European stakeholders to strengthen and expand their professional service capabilities in the HPC and AI sectors, building on the expertise and collaboration developed throughout the project lifecycle.

\textbf{Acknowledgment:}
This work has received funding from the European High-Performance Computing Joint Undertaking (JU) under grant agreement No 101175702.

\bibliographystyle{IEEEtran}
\bibliography{biblio}

\end{document}